\documentclass{JHEP3} 
\keywords{Space-Time Symmetries, Beyond Standard Model}

\usepackage{bm,subequation}

\newcommand{\xbf}{\bm{x}}
\newcommand{\pbf}{\bm{p}}
\newcommand\U{\mathop{\rm {}U}\nolimits} 

\title{Quantum theory of noncommutative fields}

\author{Jos\'e Manuel Carmona and Jos\'e Luis Cort\'es\\ 
Departamento de F\'{\i}sica Te\'orica, Universidad de Zaragoza\\ 
Zaragoza 50009, Spain\\ 
E-mail: \email{jcarmona@posta.unizar.es}, \email{cortes@posta.unizar.es}}

\author{Jorge Gamboa and Fernando M\'endez\\ 
Departamento de F\'{\i}sica, Universidad de Santiago de Chile\\ 
Casilla 307, Santiago 2, Chile\\
E-mail: \email{jgamboa@lauca.usach.cl}, \email{fmendez@lauca.usach.cl}}

\abstract{Generalizing the noncommutative harmonic oscillator
construction, we propose a new extension of quantum field theory based
on the concept of ``noncommutative fields''. Our description permits
to break the usual particle-antiparticle degeneracy at the dispersion
relation level and introduces naturally an ultraviolet and an infrared
cutoff. Phenomenological bounds for these new energy scales are
given.}

\begin{document}

\section{Introduction}

Relativistic quantum field theory (RQFT) is the general framework of our 
present microscopical theories. Its validity in particle physics has been 
confirmed by experiments covering a very wide range of energies, from the 
eV to the TeV~\cite{cutoffs}. From a modern perspective~\cite{weinberg}, RQFT 
is the necessary form adopted by any low-energy or ``effective'' theory 
satisfying  the following three principles: special relativity, quantum 
mechanics and the cluster decomposition principle, which basically states that
distant experiments yield uncorrelated results. The success of RQFT
is then a confirmation of the validity of its ingredients, which describe
correctly low-energy phenomena.

There seems however to be strong difficulties in obtaining a RQFT
containing gravitation. Indeed there is no \emph{a priori} reason why
RQFT should be the correct framework of a high-energy theory, and even
one or more of its ingredients could fail at these energies. In
particular, Lorentz invariance could not be an exact symmetry at high
energies, as recent developments in quantum gravity
suggest~\cite{amelino,qugr}. It is clear however that any high-energy
theory of particle physics should reduce to a RQFT at low energies.

In this paper we propose an extension of the RQFT framework based on
the notion of what we will call ``a noncommutative field''. Motivated
by the appearance of noncommutative spaces in string
theory~\cite{string}, there has recently been quite a few developments
on ``noncommutative quantum mechanics'' (NCQM), which is an extension
of quantum mechanics (QM) consisting in the formulation of quantum
mechanical systems on a noncommutative coordinate space (or even
phase-space)~\cite{NCQM,nair}.  ``Noncommutative quantum field
theories'', meaning quantum field theories on such spaces, have also
been studied~\cite{NCQFT}. These theories violate relativistic
invariance~\cite{carroll} and modify in a peculiar way the
short-distance behavior of the theory (although there are still
ultraviolet divergences).

There is however another way of introducing a noncommutativity in
quantum field theory. In NCQM, the coordinates (and momenta), which
are the degrees of freedom of the system, are made noncommutative.
The degrees of freedom in field theory are the fields at every point
of space (and their conjugated momenta). Therefore the natural
generalization of NCQM to field theory leads to a noncommutative field
(instead of a field in a noncommutative space). In
section~\ref{sec:NCF} we show how a quantum theory of such a field can
be constructed and then in section~\ref{sec:prop} we will study the
properties of the extension of RQFT it provides. Finally in
section~\ref{sec:phen} we will consider its phenomenological
implications, the bounds that present experimental status produce on
the size of the noncommutativity and the possibility to measure
effects coming from this source.

\section{The noncommutative field}\label{sec:NCF}

\subsection{Definition of the free noncommutative field theory}

Let us consider a scalar field theory with two fields
$(\phi_1,\phi_2)$, i.e., a complex field
$\Phi=(\phi_1+i\phi_2)/\sqrt{2}$, and let us introduce a
noncommutativity between the two fields at every point of space (for
now we work in the Schr\"odinger picture where fields and momenta do
not depend on time)
\begin{equation}
\left[\Phi(\xbf),\Phi^\dagger(\xbf')\right]=\theta \,\delta^3(\xbf-\xbf')\,.
\label{theta}
\end{equation}
We may consider at the same time a noncommutativity in the momenta
$\Pi=(\pi_1+i \pi_2)/\sqrt{2}$
\begin{equation}
\left[\Pi(\xbf),\Pi^\dagger(\xbf')\right]=B \,\delta^3(\xbf-\xbf')\,,
\label{B}
\end{equation}
where $\theta$ and $B$ are the parameters which parametrize the
noncommutativity. The fields and their conjugated momenta are related
by the conventional commutation relations
\begin{subequations}
\begin{eqnarray}
\left[\Phi(\xbf),\Pi^\dagger(\xbf')\right] &=& i \,\delta^3(\xbf-\xbf')\,, \\
\left[\Phi^\dagger(\xbf),\Pi(\xbf')\right] &=& i \,\delta^3(\xbf-\xbf')\,.
\end{eqnarray}
\label{phipi}
\end{subequations}
The free theory of the complex noncommutative quantum field is
completely defined by the above commutation relations, together with
the hamiltonian
\begin{eqnarray}
H &=& \int d^3 \xbf \,\mathcal{H}(\xbf)\,, 
\nonumber \\
\mathcal{H}(\xbf) &=&
\Pi^{\dagger}({\xbf}) \Pi({\xbf}) + 
\lambda\bm{\nabla}\Phi^{\dagger}({\xbf}) \bm{\nabla}\Phi({\xbf}) +
m^2 \Phi^{\dagger}({\xbf}) \Phi({\xbf})\,, 
\label{H}
\end{eqnarray}
where $\mathcal{H}$ is the hamiltonian density. Note that there is one
additional dimensionless parameter $\lambda$ in the hamiltonian as
compared to the canonical theory because we already set the scale of
the fields through the commutation relations~(\ref{phipi}).

The commutation relations~(\ref{theta}) and~(\ref{B}) are not the most
general ones to define a noncommutative field: one could introduce
nonzero commutators between fields (and momenta) at different spatial
points, which would then involve a much more complicated and arbitrary
parametrization than that of eqs.~(\ref{theta}) and~(\ref{B}). We will
examine the properties and characteristic features of this simple
definition for the noncommutative field in section~\ref{sec:prop}.

\pagebreak[3] 

In order to solve this theory, it is convenient to recall the method
followed in standard RQFT ($\theta=B=0$, $\lambda=1$).  There one
identifies the ``field'' as a superposition of an infinite number of
decoupled one-dimensional quantum harmonic oscillators, each with a
frequency $\omega(\pbf)=\sqrt{\pbf^2+m^2}$, and then uses the solution
of the harmonic oscillator in QM to define the Fock space of
particle-states where the field acts. This suggests that considering
the noncommutative generalization of the harmonic oscillator in QM
might help us to define a Fock space for the noncommutative field.

\subsection{Harmonic oscillator in noncommutative quantum mechanics}

Let us consider a particle in two noncommuting spatial dimensions (the
coordinates will be in correspondence with the two noncommuting real
fields) in the presence of the harmonic oscillator potential (in
correspondence with the relation between a free field theory and a
superposition of oscillators) and a constant magnetic field [in
correspondence with the noncommutativity in momenta eq.~(\ref{B})].

This system is defined by the hamiltonian
\begin{equation}
H =  \frac{\omega}{2} \left({\hat p}_1^2 + {\hat p}_2^2 + 
{\hat q}_1^2 + {\hat q}_2^2\right)
\label{Hqp}
\end{equation}
\emph{and} the commutation rules
\begin{equation}
[{\hat q}_1,{\hat q}_2]=i\,{\hat \theta}\,, \qquad
[{\hat p}_1,{\hat p}_2]=i\,{\hat B}\,, \qquad
[{\hat q}_i,{\hat p}_j]=i\,\delta_{ij}\, .
\label{[qp]}
\end{equation}
Note that we have expressed the hamiltonian and the commutation rules
in terms of adimensional phase-space coordinates (hence the small
angles over them) appropriately rescaled, and omitted $\hbar$ factors.

This problem was recently considered and solved in ref.~\cite{nair}.
Since we are interested in the case where the noncommutativity is
going to be a small correction to RQFT, we will restrict in the
following to the $\hat B\hat\theta<1$ case.  The appropriate way to
solve this problem is to identify a linear transformation of the
phase-space coordinates
\begin{equation}
{{\hat q}_1, {\hat p}_1, {\hat q}_2, {\hat p}_2} \longrightarrow
{{\hat Q}_1, {\hat P}_1, {\hat Q}_2, {\hat P}_2} \,, 
\label{linear}
\end{equation}
so that the commutation rules in the new variables are
\begin{equation} 
[{\hat Q}_1,{\hat Q}_2] =[{\hat P}_1,{\hat P}_2] = 0\,, 
\qquad
[{\hat Q}_i,{\hat P}_j]=i\,\delta_{ij}\,, 
\label{[QP]}
\end{equation} 
and the hamiltonian in the new variables is still diagonal
\begin{equation}
H =  \frac{\omega_1}{2} \left({\hat P}_1^2 + {\hat Q}_1^2\right) + 
\frac{\omega_2}{2} \left({\hat P}_2^2 + {\hat Q}_2^2\right).
\label{HQP}
\end{equation}

However, there is no a unique linear transformation~(\ref{linear})
satisfying eqs.~(\ref{[QP]}) and~(\ref{HQP}). In fact the authors of
ref.~\cite{nair} got a very special (and, after having obtained the
general result, complicated) linear transformation. In order to
adequately solve our original problem (the free theory of the
noncommutative field) it will prove convenient to work out the most
general solution of this associated quantum mechanical problem. We
will give now the final result.

This system is equivalent to a set of two decoupled one-dimensional
oscillators of frequencies\footnote{This result can also be obtained
directly from the second order equations for the operators ${\hat
q}_i$ which one gets after eliminating ${\hat p}_i$ in the Hamilton
equations $i\mathrm{d}\mathcal{O}/\mathrm{d}t=[\mathcal{O},H]$, where
$\mathcal{O}$ is a phase-space variable.}
\begin{subequations}
\begin{eqnarray}
\omega_1 &=& \omega  \left[\sqrt{ 1 + 
\left(\frac{{\hat B} - {\hat \theta}}{2}\right)^2} +
\left(\frac{{\hat B} + {\hat \theta}}{2}\right)\right] ,
\label{omega1} \\
\omega_2 &=& \omega  \left[\sqrt{ 1 + 
\left(\frac{{\hat B} - {\hat \theta}}{2}\right)^2} -
\left(\frac{{\hat B} + {\hat \theta}}{2}\right)\right] .
\label{omega2}
\end{eqnarray}
\label{omegas}
\end{subequations}
If we set ${\hat\theta}={\hat B}=0$ then $\omega_1=\omega_2=\omega$
and we recover the result of the symmetric bidimensional harmonic
oscillator of frequency $\omega$.

In order to have a simple expression for the most general linear
transformation~(\ref{linear}) which passes from the
hamiltonian~(\ref{Hqp}) and the commutation rules~(\ref{[qp]}) to the
hamiltonian~(\ref{HQP}) and the commutation rules~(\ref{[QP]}) it is
convenient to use the following combination of variables:
\begin{eqnarray}
z &=&\frac{{\hat q}_1 + i {\hat q}_2}{\sqrt{2}}\, , \qquad
w = \frac{{\hat p}_1 + i {\hat p}_2}{\sqrt{2}}\, , 
\nonumber \\
{\bar z} &=& \frac{{\hat q}_1 - i {\hat q}_2}{\sqrt{2}}\, , \qquad
{\bar w} = \frac{{\hat p}_1 - i {\hat p}_2}{\sqrt{2}}\, ,  
\label{zw}
\end{eqnarray} 
instead of the original variables ${\hat q}_1$, ${\hat p}_1$, ${\hat
q}_2$, ${\hat p}_2$. The creation-annihilation operators of the
one-dimensional oscillators of frequencies $\omega_1$, $\omega_2$ are
\begin{eqnarray}
a &=& \frac{{\hat Q}_1 + i {\hat P}_1}{\sqrt{2}}\, , \qquad
b = \frac{{\hat Q}_2 + i {\hat P}_2}{\sqrt{2}}\, , 
\nonumber \\
a^{\dagger}&=& \frac{{\hat Q}_1 - i {\hat P}_1}{\sqrt{2}}\, , \qquad
b^{\dagger}= \frac{{\hat Q}_2 - i {\hat P}_2}{\sqrt{2}}\, .
\label{ab}
\end{eqnarray}
Then the most general linear transformation that allows to solve the
problem of noncommutative quantum mechanics is
\begin{subequations}
\begin{eqnarray}
z &=& \eta \epsilon_1 \,e^{i\alpha} a +
\epsilon_2 \,e^{i\beta} b^{\dagger} \,, \\
{\bar z} &=& \eta \epsilon_1 \,e^{-i\alpha} a^{\dagger} + 
\epsilon_2 \,e^{-i\beta} b \,, \\
w &=& -i \epsilon_1 \,e^{i\alpha} a +
i \eta \epsilon_2 \,e^{i\beta} b^{\dagger}\,, \\
{\bar w}&=&i \epsilon_1 \,e^{-i\alpha} a^{\dagger}-
i \eta \epsilon_2 \,e^{-i\beta} b \,,
\end{eqnarray}
\label{zwab}
\end{subequations}
where $\alpha$, $\beta$ are two angles which parametrize the most
general linear transformation. Both angles appear only in exponential
factors accompanying the $a$ and $b$ operators. Then we can use the
freedom in the phase choice of the particle states to take, without
any lost of generality, $\alpha = \beta = 0$ in eq.~(\ref{zwab}). The
coefficients $\eta$, $\epsilon_1$ and $\epsilon_2$ are expressed in
terms of the noncommutative parameters
\begin{subequations}
\begin{eqnarray}
\eta &=& \sqrt{ 1 + \left(\frac{{\hat B} - {\hat \theta}}{2}\right)^2} 
-\left(\frac{{\hat B} - {\hat \theta}}{2}\right), \\
\epsilon_1^2 &=& \frac{{\hat B} + \eta}{1 + \eta^2} 
             = \frac{1/\eta +{\hat \theta}}{1 + \eta^2}\,, \\
\epsilon_2^2 &=& \frac{\eta -{\hat \theta}}{1 + \eta^2} 
             = \frac{1/\eta -{\hat B}}{1 + \eta^2}\,.
\end{eqnarray}
\label{etaepsilon}
\end{subequations}
Since there are only two parameters for the noncommutativity, we have
a relation between these three coefficients
\begin{equation}
\epsilon_1^2 + \epsilon_2^2 = \frac{1}{\eta} \, .
\end{equation}
Finally, we can obtain from eq.~(\ref{etaepsilon}) the following
simple relations:
\begin{equation}
\epsilon_1^2 - \eta^2 \epsilon_2^2 ={\hat B}\,, \qquad
\eta^2 \epsilon_1^2 - \epsilon_2^2 ={\hat \theta}\,.
\end{equation}

\subsection{Construction of the noncommutative field}

The noncommutative field is constructed from the above solution of the
noncommutative quantum mechanics problem by considering an oscillator
for each value of the momentum $\pbf$ of frequency
$\omega(\pbf)=\sqrt{\lambda\pbf^2+m^2}$. Then, the extension to the
complex field $\Phi$ of the expression~(\ref{zwab}) of the $z$
coordinate as a function of the creation and annihilation operators is
\begin{equation}
\Phi ({\xbf}) = \int \frac{d^{3}{\pbf}}{(2\pi)^3} 
\frac{1}{\sqrt{ \omega(\pbf)}} \left[\eta(\pbf) \epsilon_1(\pbf) \, a_{\pbf} \,
e^{i{\pbf}\cdot{\xbf}}  
+ \epsilon_2(\pbf) \, b^{\dagger}_{\pbf} \, 
e^{- i{\pbf}\cdot{\xbf}}\right],
\label{Phi}
\end{equation}
which generalizes the conventional expression of the field in RQFT as
a superposition of infinite oscillators (each for every momentum) at
every space point. Eq.~(\ref{Phi}) includes the plane-wave factors
$e^{i\pbf\cdot\xbf}$, an explicit momentum-dependence in the
coefficients $\eta$, $\epsilon_1$ and $\epsilon_2$, and a global
factor $1/\sqrt{\omega}$ to take into account the rescaling between
the adimensional coordinates and the field. The momentum has the
analogue extension of the expression of $w$ in eq.~(\ref{zwab})
\begin{equation}
\Pi ({\xbf}) = \int \frac{d^{3}{\pbf}}{(2\pi)^3} 
\sqrt{\omega(\pbf)} \left[- i\,\epsilon_1(\pbf) \, a_{\pbf} \, 
e^{i{\pbf}\cdot{\xbf}} + 
i\,\eta(\pbf) \epsilon_2(\pbf) \, b^{\dagger}_{\pbf} \, 
e^{- i{\pbf}\cdot{\xbf}}\right].
\label{Pi}
\end{equation}

Eqs.~(\ref{Phi}) and~(\ref{Pi}) (and their respective conjugated
expressions) give the fields and momenta as a function of creation and
annihilation operators. If these satisfy the commutation rules
\begin{subequations}
\begin{eqnarray}
\left[a_{\pbf}, a^{\dagger}_{\pbf'}\right] 
&=& (2\pi)^3 \delta^{3}({\pbf}- {\pbf'})\,, \\
\left [b_{\pbf}, b^{\dagger}_{\pbf'}\right] 
&=& (2\pi)^3 \delta^{3}({\pbf}- {\pbf'})\,,
\label{[ab]}
\end{eqnarray}
\end{subequations}
and we choose the noncommutative parameters of the system of quantum
mechanics corresponding to each momentum as
\begin{equation}
{\hat \theta} (\pbf) = \theta \,\omega(\pbf)\,, \qquad
{\hat B} (\pbf) = \frac{B}{\omega(\pbf)}\, ,
\label{thetaB(p)}
\end{equation}
\looseness=-1 with $\theta$, $B$ constants which do not depend of the
momentum, then the calculation of the commutators of fields and
momenta give the commutation rules~(\ref{theta}),~(\ref{B})
and~(\ref{phipi}).  This proves that the field constructed as in
eq.~(\ref{Phi}) is a representation of the noncommutative field in the
Fock space defined by the creation and annihilation operators
$a_{\pbf}$, $a^\dagger_{\pbf}$, $b_{\pbf}$ and $b^\dagger_{\pbf}$.

Moreover using the representation of fields and momenta as a linear
combination of creation and annihilation operators, eqs.~(\ref{Phi})
and~(\ref{Pi}), one can express the hamiltonian~(\ref{H}) in the form
\begin{equation}
H = \int \frac{d^{3}{\pbf}}{(2\pi)^3} \left[E_1(\pbf) \left(
a^{\dagger}_{\pbf} a_{\pbf} +\frac{1}{2}\right)
+ E_2(\pbf) \left(b^{\dagger}_{\pbf} b_{\pbf} 
+\frac{1}{2}\right)\right],
\label{Hab}
\end{equation}
which shows that the theory of the free complex noncommutative field
is a theory of free particles of two types. $E_1(\pbf)$ and
$E_2(\pbf)$ give the expressions of the energy of one of these
particles with momentum $\pbf$.  These energies are simply the
frequencies in eq.~(\ref{omegas}) of the two decoupled oscillators
appearing in the solution of the quantum mechanical system
corresponding to each momentum $\pbf$ with the parameters of
noncommutativity given in eq.~(\ref{thetaB(p)})
\begin{subequations}
\begin{eqnarray}
E_1(\pbf) &=& \omega(\pbf) \left[\sqrt{1 + \frac{1}{4} 
\left(\frac{B}{\omega(\pbf)} - \theta \omega(\pbf)\right)^2}
+ \frac{1}{2}
\left(\frac{B}{\omega(\pbf)} + \theta \omega(\pbf)\right) \right], 
\label{E12p}\\
E_2(\pbf) &=& \omega(\pbf)  \left[\sqrt{ 1 + \frac{1}{4} 
\left(\frac{B}{\omega(\pbf)} - \theta \omega(\pbf)\right)^2} 
- \frac{1}{2}
\left(\frac{B}{\omega(\pbf)} + \theta \omega(\pbf)\right) \right].
\label{E12a}
\end{eqnarray}
\label{E12}
\end{subequations}

In summary we have seen that the free theory of the scalar
noncommutative field can be solved in a similar way as in the
conventional case of RQFT. It is a theory of free particles. The
simplest way to incorporate interactions is by using the same
hamiltonians as in RQFT, now in terms of noncommutative fields. Then
the solution of the free theory can be taken as a starting point for a
perturbative treatment of interactions analogously to what is done in
RQFT: identification of propagators, Feynman rules, etc. We will
sketch this procedure in the following section, and examine the
characteristic properties of the noncommutative theories defined in
such a way.

\section{Properties of the quantum theory of noncommutative fields}
\label{sec:prop}

The free noncommutative field was defined by the commutation
relations~(\ref{theta}),~(\ref{B}) and by the hamiltonian~(\ref{H}).
This simple implementation of a noncommutativity in field space has
the following properties:
\begin{enumerate} 
\item Standard RQFT is trivially recovered in the $\theta\to 0$, $B\to
0$, $\lambda\to 1$ limit. As in the standard case, the quantum theory
is obtained from a classical hamiltonian which is relativistic
invariant [the apparent noninvariance of $\lambda\neq 1$ in
eq.~(\ref{H}) is fictitious; as we remarked before, it is a
consequence of the choice of the scale for the fields in order to
write the standard commutation relations between fields and momenta
eqs.~(\ref{phipi})], but now we follow a quantification procedure,
given by the new commutation relations~(\ref{theta}) and~(\ref{B}),
which explicitely violates Lorentz symmetry. This is what we
understand by the ``quantum theory of a noncommutative field''.
\item The hamiltonian density defined in eq.~(\ref{H}) is made of
fields which commute at different space points, and therefore
satisfies
\begin{equation}
[\mathcal{H}(\xbf),\mathcal{H}(\xbf')]=0\qquad \mathrm{for}\quad
\xbf\neq\xbf'\,.
\label{density1}
\end{equation}
This property is essential in RQFT to guarantee that the $S$-matrix
will be Lorentz-invariant. More specifically, what is required in RQFT
is that
\begin{equation}
[\mathcal{H}(x),\mathcal{H}(x')]=0\qquad \mathrm{for}\quad
(x-x')^2\geq 0\,,
\label{density2}
\end{equation}
which is equivalent to eq.~(\ref{density1}) in a Lorentz-invariant
theory.  The two conditions are not equivalent however when
relativistic invariance is lost. In fact, in the quantum theory of the
noncommutative field, the hamiltonian density satisfies
eq.~(\ref{density1}) but not eq.~(\ref{density2}). The preservation of
the property~(\ref{density1}) is in any case welcome since it allows
to speak consistently about the concept of a hamiltonian
density. Without this condition, the energy of a closed finite system
could depend on the energy of another system very far away.
Commutation relations more general than eqs.~(\ref{theta})
and~(\ref{B}) would violate eq.~(\ref{density1}).
\item Keeping the property~(\ref{density1}) requires the introduction of
two real fields as the only way to implement a noncommutativity in field space.
This leads to a theory with two types of particles which correspond to
the particle and the antiparticle in the $\theta, B\to 0, \lambda\to 1$ 
limit (conventional RQFT). 
Particle and antiparticle are no longer degenerated in this extension
of RQFT, and their energy is different from the standard expression 
$\sqrt{\pbf^2+m^2}$ by small corrections parametrized by $\theta$, $B$
and $\lambda$. The theory naturally incorporates in this way a 
matter-antimatter asymmetry.
\item We have a new, specific form of the dispersion relation, or
relation between energy and momentum of a particle, eq.~(\ref{E12}),
which is no longer Lorentz-invariant, while the theory still preserves
rotational symmetry. Relativistic invariance is therefore an
ingredient which is lost in this extension of RQFT. We will see in
section~\ref{sec:phen} that this symmetry is violated not only at high
energies, but, surprisingly enough, also at low energies, being still
compatible with phenomenological observations. Relativistic causality
is also violated, as we check later in this section.
\item An essential property of RQFT, which in principle should hold in
any sensible physical theory, is the cluster decomposition principle:
experiments which are sufficiently separated in space should have
unrelated results.  A general theorem states that the $S$-matrix
satisfies this crucial requirement if the hamiltonian can be expressed
as a sum of products of creation and annihilation operators, with
suitable non-singular coefficients~\cite{weinberg}. This theorem
garantees that the cluster property still holds in the noncommutative
extension of RQFT.
\end{enumerate}

To discuss causality and the formulation of perturbation theory we
have to consider the field operator in the interaction picture
\begin{eqnarray}
\Phi ({\xbf}, t) &=& e^{i H_0 t}  \,\Phi ({\xbf})\, e^{- i H_0 t} 
\nonumber \\
&=& \int \frac{d^{3}{\pbf}}{(2\pi)^3} \frac{1}{\sqrt{\omega(\pbf)}}  
\left[\eta(\pbf) \epsilon_1(\pbf) \, a_{\pbf} \, 
e^{-i E_1(\pbf) t} e^{i{\pbf}.{\xbf}}
+ \epsilon_2(\pbf) \, b^{\dagger}_{\pbf} \, e^{i E_2(\pbf) t} 
e^{- i{\pbf}.{\xbf}}\right].
\label{Phi(t)}
\end{eqnarray}

From eq.~(\ref{Phi(t)}) one can calculate the commutator of operators
at different times
\begin{equation}
\left[\Phi(\xbf, t) , \Phi^{\dagger}\left(\xbf', t'\right)\right]
\label{[t]}
\end{equation}
and verify that it is different from zero at causally disconnected
points ($(\xbf-\xbf')^2>(t-t')^2$) owing to the noncommutativity.

The modification to the standard propagator caused by the
noncommutativity is rather simple. One has
\begin{equation} 
\langle 0|{\cal T}(\Phi(\xbf,t) \Phi^{\dagger}(\xbf', t')|0\rangle =
\int \frac{d^4 p}{(2\pi)^4} \,e^{-i p^0 (t-t')}
e^{i {\pbf}\cdot(\xbf-\xbf')} 
\frac{i \,(1 - \theta B + \theta p^0)}
{(p^0 {-} E_1(\pbf) {+}i\epsilon) (p^0 {+} E_2(\pbf) {-}i\epsilon)}\, , 
\label{propagador}
\end{equation}
that is, the effect of the noncommutativity is a displacement in the
position of the poles
\begin{equation}
\pm\sqrt{\pbf^2+m^2}\,\to E_1(\pbf), -E_2(\pbf)\,, 
\end{equation}
together with a modification of the residues
\begin{equation}
\pm \frac{1}{2\sqrt{\pbf^2+m^2}}\to 
\frac{(1-\theta B+\theta E_1(\pbf))}{(E_1(\pbf)+E_2(\pbf))},
-\frac{(1-\theta B-\theta E_2(\pbf))}{(E_1(\pbf)+E_2(\pbf))}\,.
\end{equation}

\looseness=-1 We will finally note that there is not any obstruction to the
introduction of gauge symmetries in the theory of noncommutative
fields. In the free noncommutative complex scalar field theory we have
a global $\U(1)$ symmetry that can be made local in a hamiltonian with
interaction terms containing so many $\Phi$ fields as $\Phi^\dagger$
fields, if every derivative of the field appears in the combination
$-i\,\bm{\nabla} \Phi + \bm{A}\Phi$. We leave further exploration on
the dynamics of interactions in theories of noncommutative fields for
future work and consider in the following section the phenomenological
implications coming from the solution of the free theory.
   
\section{Phenomenological bounds on the parameters of noncommutativity}
\label{sec:phen}

We would like to show in this section how the quantum theory of
noncommutative fields can be a sensible extension of RQFT, in the
sense that it does not contradict the present understanding of
low-energy phenomena and, at the same time, may have observable
consequences.

The effect of the noncommutativity at the level of the free theory is
the substitution of the particle or antiparticle states of momentum
$\pbf$ and energy $E=\sqrt{\pbf^2+m^2}$ in RQFT by two states of
energies $E_1(\pbf)$ and $E_2(\pbf)$ given by eqs.~(\ref{E12}). From
these expressions one sees that if
\begin{equation}
B \ll \sqrt{\lambda\pbf^2 + m^2} \,\ll\, 1/\theta 
\Rightarrow E_1(\pbf) \approx E_2(\pbf) \approx   
\sqrt{\lambda\pbf^2 + m^2}
\end{equation}
and the relativistic dispersion relation is recovered in the
$\lambda=1$ limit.

\pagebreak[3] 

We then have two energy scales coming from the
noncommutativity: $B$ and $1/\theta$, and from the dispersion relation
we see that these two scales are respectively the infrared (IR) and
ultraviolet (UV) scales which limit the range of validity of the
relativistic invariant theory.

The noncommutativity induces a violation of relativistic invariance
both at high and low energies.  A maximum velocity of propagation
different from the speed of light $(\lambda\neq 1)$ and high-energy
violations coming from the presence of an UV scale are properties
which have been explored in other
contexts~\cite{cutoffs,amelino,qugr,kostel,uv}, especially those
trying to incorporate effects coming from the Planck length. What is
new in the present extension of RQFT is the presence of an additional
IR scale and a violation of relativistic invariance at low~energies.

This result allows to explore consequences of the noncommutativity
already at the level of the kinematics, without a necessity of
considering in detail the dynamics. The analysis will alternatively
give restrictions on the values of the parameters of the
noncommutativity, based on the success of the RQFT description of
nature. We note here that this kinematics study needs a further
assumption. The new dispersion relations~(\ref{E12}) were obtained for
the free theory of the scalar noncommutative field. The extension to a
free theory of fermions is not however a trivial task, and we leave it
for future work.  We will now make the phenomenological analysis to
obtain bounds on the parameters of noncommutativity assuming that a
similar dispersion relation holds for fermions.

The most sensitive experiments to detect a violation of relativistic
invariance at high energies are those involving ultra high-energy
cosmic rays (UHECR).  Not only they reach energies as high as
$10^{20}$ eV, but they are even sensitive to effects parametrized by
much larger energy scales, such as the Planck scale, thanks to
amplification mechanisms coming from the presence of very different
scales. This is what happens at the very end of the cosmic ray
spectrum: relativistic kinematics predicts a cutoff in the spectrum
for UHECR coming from distant sources (the GZK cutoff)~\cite{gzk},
caused by the energy loss they experiment in their interaction with
the cosmic background radiation (CBR), which seems to be avoided
somehow~\cite{gzkexps}.  Relativistic invariance violations induced by
the Planck scale are, among others~\cite{othersgzk}, a possible
explanation for the disappearance of the GZK
cutoff~\cite{amelino,uv,capolavoro,grillo}.  The sensitivity to the
Plank scale results in this case from the presence of a very small
energy scale: the kinetic energy of photons of the CBR, $10^{-3}$ eV.

In this range of momenta, we expect $|\theta|\,(\pbf^2+m^2)\gg |B|$,
and then $B$ can be neglected in the dispersion
relations~(\ref{E12}). Bounds on $1/\theta$ in the $B\to 0$ limit
coming from the physics of the UHECR were studied in
ref.~\cite{capolavoro}. Taking the energy-momentum relation
eq.~(\ref{E12p}) for the particle, and eq.~(\ref{E12a}) for the
antiparticle, so that the noncommutative field eq.~(\ref{Phi})
generalizes the expression of the conventional field in RQFT, as a
linear combination of particle annihilation operators $a_{\pbf}$ and
antiparticle creation operators $b^\dagger_{\pbf}$, then the analysis
of ref.~\cite{capolavoro} shows that the sign of $\theta$ has to be
negative.  This is because $E_1(\pbf)$ with $\theta>0$ would generate
a mechanism of energy loss for particles independent of the CBR:
particle disintegrations prohibited by relativistic kinematics, would
be now allowed.  The observation of UHECR excludes then this
possibility. With $\theta<0$, $E_1(\pbf)<\omega(\pbf)$, and the
interaction with the CBR is now kinematically forbidden so that the
GZK cutoff no longer exists. Experiments coming in the near
future~\cite{newgzkexps} will clarify the situation with respect to
the GZK cutoff violation, which will then be a stringent test for the
theory of noncommutative fields. The bounds provided for the UV energy
scale are~\cite{capolavoro}: $10^{21}\mbox{ eV}\lesssim 1/|\theta|
\lesssim 10^{43}\mbox{ eV}$.

Let us now consider the IR corrections to the relativistic dispersion
relation given by eqs.~(\ref{E12}). In the range of momenta where
$|\theta|\,(\pbf^2+m^2)\ll |B|$, we can neglect in a first
approximation the effects parametrized by the UV scale ($1/\theta$)
and then we obtain
\begin{equation}
E_1(\pbf)\approx \sqrt{\pbf^2+m^2+\frac{B^2}{4}}\,-\frac{B}{2}
\end{equation}
for the energy of the particle, and an analogous expression for the
antiparticle (replacing $-B/2$ by $+B/2$). We then see that the effect
of the noncommutativity is a constant contribution to the energy
(opposite in sign for particles and antiparticles) and a
``renormalization'' of the mass $m^2 \to m^2_{\mathrm{\rm
eff}}=m^2+B^2/4$. Conservation laws in physical process will however
make invisible the $\pm B/2$ constant contributions to the energy
coming from the noncommutativity. On the other hand, the bound
$B^2/4\leq m^2_{\mathrm{\rm eff}}$ will restrict the value of $B$ from
the bounds to neutrino masses. The $\beta$ disintegration of tritium,
which is the most sensitive experiment to neutrino
mass~\cite{tritioexps}, gives $B<5$ eV.

Finally, bounds on the adimensional parameter $\lambda$ were
considered in ref.~\cite{uv} in different scenarios. The typical bound
is $|1-\lambda|\leq 10^{-23}$. A more detailed phenomenological
analysis including cosmological implications of the noncommutativity
will be given elsewhere.
  
\acknowledgments
We would like to thank J. Clemente-Gallardo for discussions.  This
work has been partially supported by the grants 1010596, 7010596 and
3000005 from Fondecyt-Chile, by M.AA.EE./AECI and by MCYT (Spain),
grants FPA2000-1252 and FPA2001-1813.

\end{document}